\documentclass[conference]{IEEEtran}

\IEEEoverridecommandlockouts
\usepackage{graphicx}
\usepackage{subfiles}
\usepackage{array}
\usepackage{amsmath}
\usepackage{algorithm}
\usepackage{algpseudocode}
\usepackage{multirow}
\usepackage{subfigure}
\usepackage{caption}
\usepackage{flushend}
\usepackage{balance}
\usepackage{diagbox}
\usepackage{ulem}
\usepackage{bm}
\usepackage{amsfonts}
\usepackage{hyperref}

\usepackage{tcolorbox}

\usepackage{float}
\usepackage{enumitem}

\usepackage[framemethod=TikZ]{mdframed}
\usepackage{tcolorbox}
\makeatletter
\newcommand{\mybox}[1]{%
  \setbox0=\hbox{#1}%
  \setlength{\@tempdima}{\dimexpr\wd0+13pt}%
  \begin{tcolorbox}[boxrule=0.5pt, colback=white, arc=4pt,
      left=6pt,right=6pt,top=6pt,bottom=6pt,boxsep=0pt]
    #1
  \end{tcolorbox}
}
\usepackage{tikz}
\usepackage{pgfplots}

\usepackage{color}
\usepackage{xcolor,colortbl}
\usepackage{array}
\usepackage{amsmath}
\usepackage{centernot}
\usepackage{xspace}
\usepackage{url}
\usepackage{verbatim}
\usepackage{wrapfig}
\usepackage{tabularx}
\usepackage{cite}

\newcommand{\tool}{ISPY}

\newcommand{\numPair}{30K}
\newcommand{\numPairExp}{21K}
\newcommand{\numAns}{26}
\newcommand{\numdataset}{750}
\newcommand{\website}{https://github.com/jzySaber1996/ISPY}

\newcommand{\attribute}{15}

\definecolor{colorEN}{RGB}{255, 255, 224}
\definecolor{colorES}{RGB}{255, 246, 143}
\definecolor{colorEM}{RGB}{238, 201, 0}
\definecolor{colorEL}{RGB}{205, 149, 12}

\definecolor{CadetBlue2}{RGB}{142, 229, 238}
\definecolor{Cyan3}{RGB}{188, 238, 104}
\definecolor{Turquoise3}{RGB}{70, 130, 180}

\def\BibTeX{{\rm B\kern-.05em{\sc i\kern-.025em b}\kern-.08em
    T\kern-.1667em\lower.7ex\hbox{E}\kern-.125emX}}

\begin{document}

\title{{\tool}: Automatic Issue-Solution Pair Extraction from Community Live Chats\\
}

\author{
\IEEEauthorblockN{
Lin Shi\IEEEauthorrefmark{1}\IEEEauthorrefmark{4}\footnotemark$^1$,
Ziyou Jiang\IEEEauthorrefmark{1}\IEEEauthorrefmark{4}\footnotemark$^1$,
Ye Yang\IEEEauthorrefmark{5},
Xiao Chen\IEEEauthorrefmark{1}\IEEEauthorrefmark{4},
Yumin Zhang\IEEEauthorrefmark{1}\IEEEauthorrefmark{4},
Fangwen Mu\IEEEauthorrefmark{1}\IEEEauthorrefmark{4},\\
Hanzhi Jiang\IEEEauthorrefmark{1}\IEEEauthorrefmark{4},
Qing Wang\IEEEauthorrefmark{1}\IEEEauthorrefmark{2}\IEEEauthorrefmark{3}\IEEEauthorrefmark{4}\footnotemark$^2$,
}

\IEEEauthorblockA{\IEEEauthorrefmark{1}Laboratory for Internet Software Technologies, \IEEEauthorrefmark{2}State Key Laboratory of Computer Sciences, }

\IEEEauthorblockA{\IEEEauthorrefmark{3}Science\&Technology on Integrated Information System Laboratory,
}

\IEEEauthorblockA{Institute of Software Chinese Academy of Sciences, Beijing, China;
}

\IEEEauthorblockA{\IEEEauthorrefmark{4}University of Chinese Academy of Sciences, Beijing, China;}

\IEEEauthorblockA{\IEEEauthorrefmark{5}School of Systems and Enterprises, Stevens Institute of Technology, Hoboken, NJ, USA;}

\IEEEauthorblockA{Email: \{shilin, ziyou2019, chenxiao2021, yumin2020, fangwen2020, hanzhi2021, wq\}@iscas.ac.cn, yyang4@stevens.edu}
}




\maketitle
\footnotetext[1]{Both authors contributed equally to this research.}
\footnotetext[2]{The corresponding author.}

\begin{abstract}


Collaborative live chats are gaining popularity as a development communication tool. In community live chatting, developers are likely to post issues they encountered (\textit{e.g.}, setup issues and compile issues), and other developers respond with possible solutions. Therefore, community live chats contain rich sets of information for reported issues and their corresponding solutions, which can be quite useful for knowledge sharing and future reuse if extracted and restored in time. However, it remains challenging to accurately mine such knowledge due to the noisy nature of interleaved dialogs in live chat data.
In this paper, we first formulate the problem of issue-solution pair extraction from developer live chat data, and propose an automated approach, named {\tool}, based on natural language processing and deep learning techniques with customized enhancements, to address the problem.
Specifically, {\tool} automates three tasks:
1) Disentangle live chat logs, employing a feedforward neural network to disentangle a conversation history into separate dialogs automatically; 
2) Detect dialogs discussing issues, using a novel convolutional neural network (CNN), 
which consists of a BERT-based utterance embedding layer, a context-aware dialog embedding layer, and an output layer;
3) Extract appropriate utterances and combine them as corresponding solutions, {based on the same CNN structure but with different feeding inputs.}
To evaluate {\tool}, we compare it with six baselines, 
utilizing a dataset with {\numdataset} dialogs including 171 issue-solution pairs and evaluate {\tool} from eight open source communities.
The results show that, for issue-detection, our approach achieves the \textit{F1} of 76\%, and outperforms all baselines by 30\%.
Our approach achieves the \textit{F1} of 63\% for solution-extraction and outperforms the baselines by 20\%.
Furthermore, we apply {\tool} on three new communities to extensively evaluate {\tool}'s practical usage. Moreover, we publish over {\numPair} issue-solution pairs extracted from 11 communities.
We believe that {\tool} can facilitate community-based software development by promoting knowledge sharing and shortening the issue-resolving process. 
\end{abstract}
\section{Introduction}
Synchronous communication via live chats 
allows developers to seek information and technical support, share opinions and ideas, discuss issues, and form community development \cite{chatterjee2019exploratory, DBLP:journals/jss/ChatterjeeKP20},
in a more efficient way compared with asynchronous communication such as emails or forums \cite{DBLP:conf/cscw/LinZSS16,DBLP:conf/msr/ShihabJH09,DBLP:conf/icsm/ShihabJH09}. Consequently, live chatting has become an integral component of most software development processes, not only for open source communities 
constituting globally distributed developers, but also for software companies to facilitate in-house team communication and coordination, esp. in accommodating remote work due to the COVID-19 pandemic \cite{DBLP:journals/corr/abs-2101-05877}. 

\begin{figure}[t]
\centering
\includegraphics[width=\columnwidth]{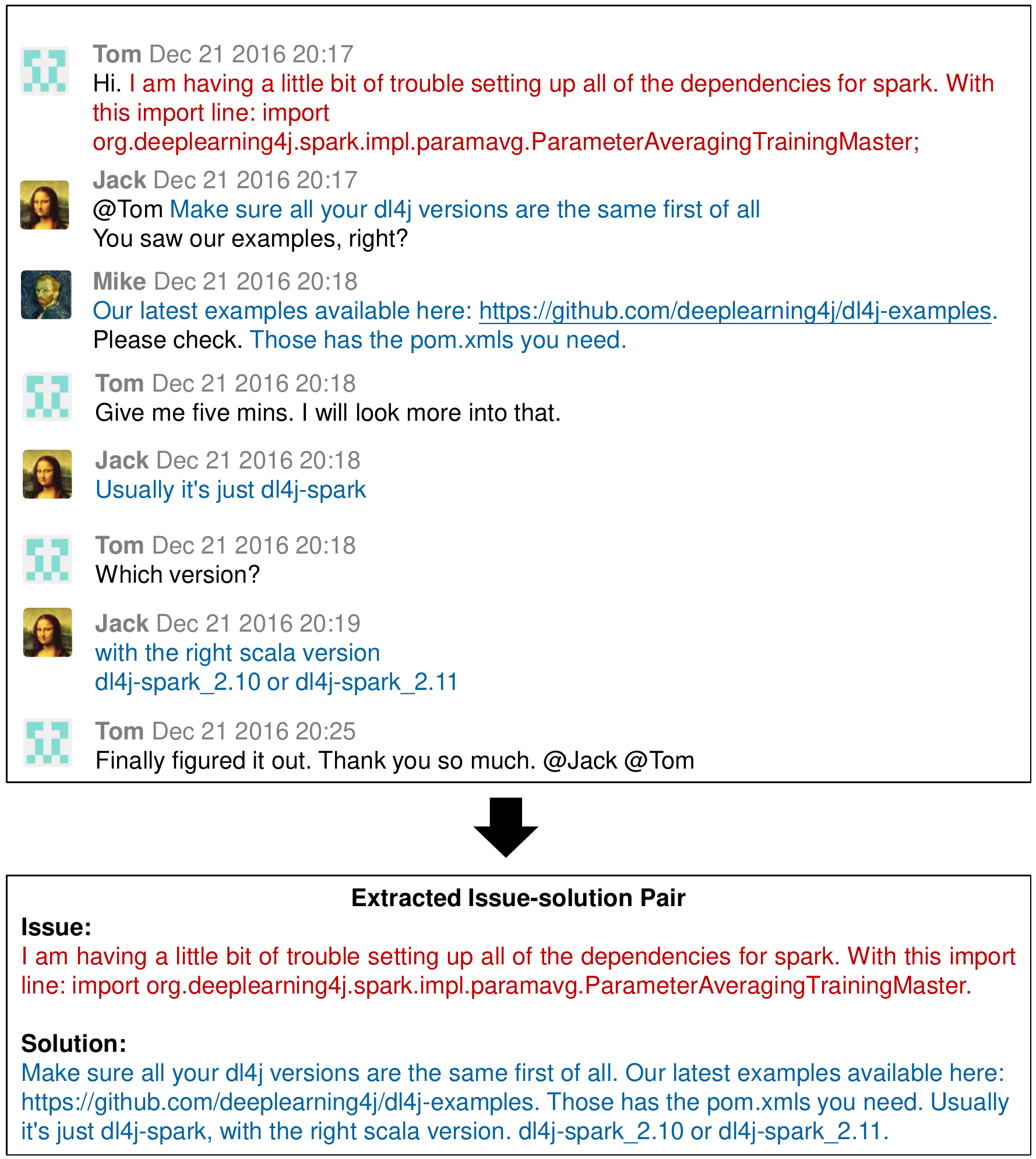}
\caption{An example of issue-solution pair extraction from the Deeplearning4j live chats.}
\label{fig:motivation}
\vspace{-0.5cm}
\end{figure}

Existing literature reports various motivation factors for live chatting practices.
{One of the frequently mentioned factors is to use live chats as a tool for issue-solving, such as 
installation and setup issues, bug resolution, build and compile issues \cite{DBLP:journals/tosem/EhsanHMZ21,shi2021look}.} In such cases, developers post questions related to
some specific issues, and rely on others to provide potential solutions. Alkadhi et al. \cite{DBLP:conf/msr/AlkadhiLGB17,DBLP:conf/wcre/AlkadhiNGB18} analyzed 8,702 chat messages of three OSS development teams, and found 24\% of the messages are reporting issues, and 51\% of the messages are proposing alternative issue solutions. 
As a result, live chat repositories usually contain rich information to shed light on knowledge regarding frequent issue-solution pairs. 
Fig. \ref{fig:motivation} illustrates an example slice of live chat data. 
In this scenario, \textit{Tom} encountered trouble when setting up dependencies for spark, so he posted an issue in live chatting. \textit{Jack} and \textit{Mike} both provided solutions as well as suggested examples. With their help, \textit{Tom} finally resolved that issue. From this conversation slice, we can extract the issue description, as highlighted in red, and alternative solutions colored in blue. 

However, 
it is quite challenging to mine issue-solution pairs from live chats due to the following barriers. 
 \textbf{(1) Entangled dialogs}. 
Live chat data gets big rapidly, and multiple concurrent discussions regarding different issues frequently exist in an interleaved manner.  
In order to perform any kind of dialog-level analysis, it is essential to have automated support for identifying and dividing sequential utterances into a set of distinct dialogs, according to the issue topics.
\textbf{(2) Expensive human effort.} 
Chat logs are typically high-volume and contain informal dialogs covering a wide range of technical and complex topics. It is necessary to leverage manual annotation to 
guide the construction and training of learning-based algorithms. However, the manual annotation process 
requires experienced analysts to spend a large amount of time so that they can understand the dialogs thoroughly. Thus, it is very expensive to classify issue-related dialogs. 
\textbf{(3) Noisy data}.
There exist noisy utterances such as duplicate and unreadable messages in the chat log that do not provide any valuable information.
The noisy data poses a difficulty to analyze and interpret the communicative dialogs. 

In this paper, we propose a novel approach, named {\tool} (extracting \textbf{I}ssue-\textbf{S}olution \textbf{P}airs from communit\textbf{Y} live chats) to automatically extract issue-solution pairs from development community live chats. 
{\tool} addresses the problem with three elaborated sub-tasks: 1) Disentangle live chat logs, employing a feedforward neural network to automatically disentangle a conversation history into separate dialogs; 2) Detect dialogs that are discussing issues, using a novel convolutional neural network (CNN), which consists of a BERT-based utterance embedding layer, a context-aware dialog embedding layer, and an output layer;
and 3) Extract appropriate utterances and combine them as corresponding solutions, {based on the same CNN structure but with different feeding inputs.}
To evaluate {\tool}, we first collect and utilize a dataset with {\numdataset} dialogs including 171 issue-solution pairs, and evaluate {\tool} from eight Gitter communities.
The results show that, for issue-detection, our approach achieves the \textit{F1} of 76\%, and outperforms baselines by 30\%.
For solution-extraction, our approach achieves the \textit{F1} of 63\%, and outperforms baselines by 20\%.
Furthermore, we apply {\tool} on three new communities to extensively evaluate its practical usage. {\tool} helps provide solutions for {\numAns} recent issues posted on Stack Overflow.
Adding up the {\numPairExp} pairs extracted from the former eight communities, we publish over {\numPair} issue-solution pairs extracted from 11 communities in total.
We believe that {\tool} can facilitate community-based software development by promoting knowledge sharing and shortening the issue-resolving process.
The major contributions of this paper are:
\begin{itemize}
\item We formulate the problem of issue-solution pair extraction from developer live chat data. To the best of our knowledge, this is the first study exploring this problem.
\item We propose an automated approach, named {\tool}, based on a convolutional neural network and introduces several customized improvements to
effectively handle the characteristics of this task. 
\item We evaluate the {\tool} by comparing with six baselines, with superior performance.
\item We open-source a replication package and a large dataset with over {\numPair} issue-solution pairs extracted by {tool} from 11 active communities 
on our website: {\url{\website}}.
\end{itemize}

In the remainder of the paper, Section II illustrates the problem definition. Section III presents the approach. Section IV sets up the experiments. Section V describes the results and analysis. Section VI illustrates the practical usage. Section VII is the discussion and threats to validity. Section VIII introduces the related work. Section IX concludes our work. 
\section{Problem Definition}
\label{sec:background}

\begin{figure*}[t]
\centering
\includegraphics[width=\textwidth]{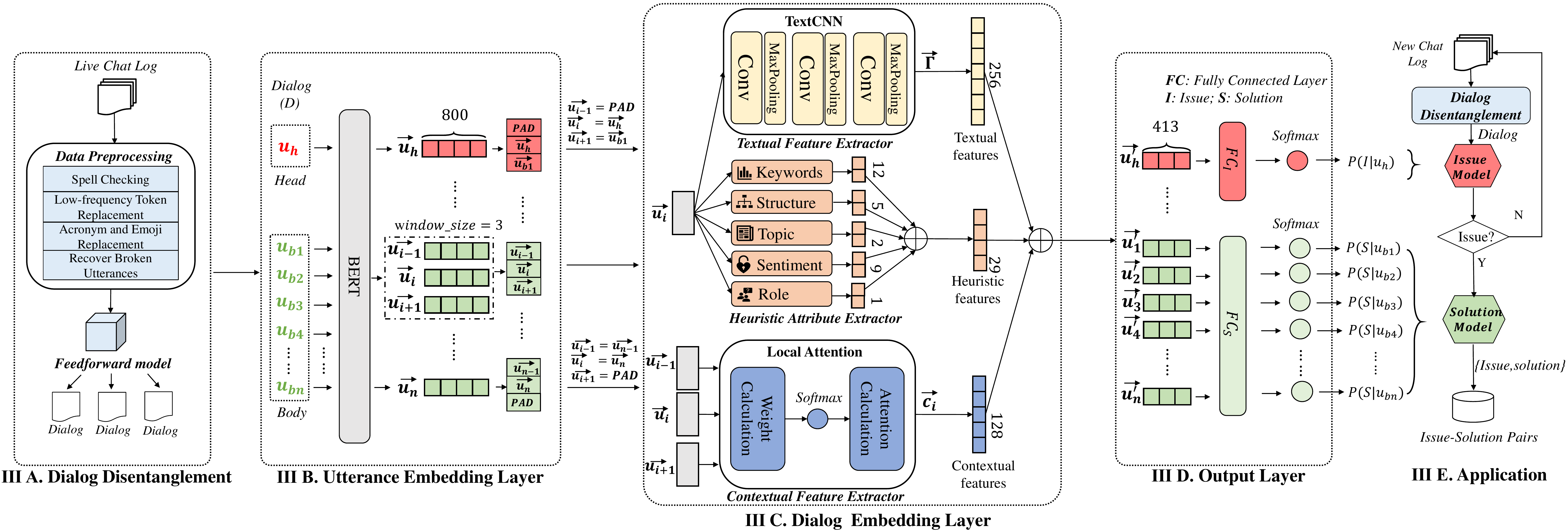}
\caption{The overview of {\tool}.}
\vspace{-0.5cm}
\label{fig:model}
\end{figure*}

Three main concepts about community live chats are concerned with this study's scope, including \textit{chat log, utterance, and dialog}. Developer conversations in one chatting room are recorded in a chat log. As illustrated in Fig. \ref{fig:motivation}, a typical live chat log contains a sequential set of utterances in chronological order.  
Each utterance consists of a timestamp, developer id, and a textual message initiating a question or responding to an earlier message. {A chat log includes all the utterances sent among participants who have been chatting in the room. Typically, it contains a large number of utterances, and the utterances might be responding to different threads of conversations. We define dialog as the conversation between two or more participants toward exploring a particular subject (e.g., resolution of a problem).}

\begin{equation}
\begin{aligned}
    &Chat\_log = \{u_1,u_2,...,u_n\}  \\
    &D = \{u_i|S(u_i)=s, u_i \in Chat\_log\} \\
    &u = <time, id, text>
\end{aligned}
\end{equation}

Equation (1) provides the definitions for the three main concepts. Specifically, a chat log $Chat\_log$ corresponds to a sequence of $n$ utterances in chronological order. A dialog $D$ is a subset of $Chat\_log$, containing only those utterances responding to the same subject $s$, {which can be determined by clustering techniques \cite{li2020dialbert,ijcai20e2e} or probability distribution estimation methods \cite{ffmodel,bilstm2017usage,pointnet}.} $S(u_i)$ denotes the subject of utterance $u_i$, and
each utterance $u$ consists of the timestamp, developer id, and textual message. 

Our work automatically targets extracting issue-solution pairs from community live chats. 
First, we divide one dialog $D$ into two parts: \textit{Head} and \textit{Body}.
\begin{equation}
\begin{aligned}
    & Head = u_h = u_{h1}\oplus u_{h2}\oplus ... \oplus u_{hm} \\
    & Body = \{u_{b1}, u_{b2}, ... , u_{bn}\} \\
    & D = \{Head, Body\} \\
\end{aligned}
\end{equation}
where \textit{Head} (also marked as $u_h$) is the concatenation of all the utterances that the dialog initiator posts before the first reply from other developers. \textit{Body} is the set of the remaining utterances. Dialog $D$ is their joint set. Based on this division, we introduce a simplification assumption that the issue descriptions appear at the head utterances authored by the dialog initiator, while solution utterances are likely to appear afterward.

Following these concepts and assumption, we formulate the problem of automatic issue-resolution pair extraction with three elaborated sub-tasks:

\begin{enumerate}
\item Dialog disentangle: Given the historical chat log $Chat\_log$, disentangle it into separate dialogs \{$D_1,D_2,...,D_n$\}.
\item Issue detection: Given a separate dialog $D_i$, find a binary function $f$ so that $f({Head}_i)$ can determine whether the dialog head depicts issue. 
\item Solution extraction: Given a dialog $D_i$ involving issue discussion, find a function $g$ 
so that $g({Body}_i)=\{u_{s1}, u_{s2},...,u_{sm}\}$, where $u_{si}$ is the utterance within the dialog suggesting potential solutions.
\end{enumerate}

Therefore, the output of our approach is a set of issue-solution pairs. Ideally, users do not need other information (e.g., the utterances between them) to understand these pairs.

\section{Approach}
The construction of {\tool} consists of four main steps, as illustrated in Fig. \ref{fig:model}. The first step includes data preprocessing and dialog disentanglement using a feedforward model. The second step is to construct an utterance embedding layer, which embeds tokenized utterances into vectors with local window context. The third step is to construct a dialog embedding layer, which defines and extracts three sets of features characterizing the context of potential issues or solutions. The fourth step is the output layer, which predicts two outputs, i.e., the possibility of issue description, and the possibility of solutions, for the corresponding inputs. 

By feeding dialog head and body into {\tool} separately, we can obtain two models: issue model and solution model. Finally, {\tool} apply the issue model and solution model for constructing pairs.

\subsection{Dialog Disentanglement}
\begin{table*}[tp]
\caption{Details about {\attribute} heuristic attributes.}
\vspace{-0.2cm}
\centering
 \resizebox{\textwidth}{!}{
\begin{tabular}{|c|c|m{9.5cm}|m{6.5cm}<{\centering}|}
\hline
\textbf{Type} & \textbf{Variable} &
\multicolumn{1}{c|}{\textbf{Description}} & \textbf{Example Value} \\
\hline
\multirow{5}{*}{Keyword}
& 5W1H & Occurrence of ``What", ``Why", ``When", ``Who", ``Which", and ``How". & \{what=1, why=0, when=0, who=0, which=0, how=0\}\\ 
\cline{2-4} 
& Punctuation & Occurrence of ``?" and ``!". & \{``?"=1, ``!"=0\}\\ 
\cline{2-4} 
& Greeting & Occurrence of greeting words: ``Hello", ``Good Morning", ``Hi Guys", etc. & False\\ 
\cline{2-4} 
& Disapproval & Occurrence of disapproval words: ``no", ``can't work", ``break down" etc. & False \\ 
\cline{2-4} 
& Mention & Occurrence of ``simi-" and ``same". & \{simi-=0, same=0\} \\ 
\hline
\multirow{5}{*}{Structure}
& NT & Number of tokens in utterance. & 7 \\ 
\cline{2-4} 
& NUT &  Number of unique tokens in utterance. & 7  \\ 
\cline{2-4} 
& NST & Number of unique tokens after stemming. & 6 \\ 
\cline{2-4} 
& AP & Absolute position of utterance in dialog. & 2  \\ 
\cline{2-4} 
& RP & Relative position of utterance in dialog. & 0.20 \\ 
\hline
\multirow{2}{*}{Topic}
& TDH & Topic deviation between the head of dialog and the entire chat. & 0.33  \\ 
\cline{2-4} 
& TDU & Topic deviation between the head of dialog and the given utterance. & 0.42   \\ 
\hline
\multirow{3}{*}{Sentiment}
& SS & Polarity sentiment scores from NLTK \cite{nltk.org} for positive, intermediate, and negative. & \{pos.=0.25, int.=0.31, neg.=0.70\}\\ 
\cline{2-4} 
& SW & Number of three-type polarity sentiment words (pos., int., neg.). & \{pos.=1, int.=5, neg.=1\}\\ 
\cline{2-4} 
& SE  & Number of three-type polarity sentiment emojis (pos., int., neg.). & \{pos.=0, int.=0, neg.=1\}\\ 
\hline
Role
& DI & Whether participant is the dialog initiator. & False \\
\hline
\end{tabular}}
\vspace{-0.5cm}

\label{tab:heurustics}
\end{table*}

\subsubsection{Data Preprocessing}
For data preprocessing, we first follow the standard pipeline of stopword removal, typo correction, lowercase conversion, and lemmatization with Spacy \cite{spacy.io}.
Additionally, due to the unique characteristics of live chat data, we employ additional data preprocessing techniques to handle special issues. Specifically, we first replace low-frequency tokens such as URL, email address, code, HTML tag, and version number with specific tokens \textit{[URL], [EMAIL], [HTML], [CODE]} and \textit{[ID]} respectively. Second,
we replace the acronym words with their full names by referring to the Oxford abbreviation library \cite{public.oed.com}. 
Following previous work \cite{DBLP:journals/chb/BoutetLCC21}\cite{DBLP:journals/cogcom/SumanSBC21}, we normalize the emojis with specific strings to standard ASCII strings. Finally, we combine consecutive utterances that are broken from one sentence according to the perplexity scores \cite{perplexity} calculated by Baidu AI Cloud \cite{intl.cloud.baidu.com}.
Following the experience from a recent study \cite{DBLP:journals/corr/JozefowiczVSSW16}, we use the perplexity scores lower than 40 as the threshold value to combine broken sentences. 

\subsubsection{Dialog Disentanglement Model}
Utterances from a single conversation thread are usually interleaved with other ongoing conversations. In this step, we focus on
dividing chat utterances into a set of distinct conversations, 
leveraging on Kummerfeld et al.'s technique \cite{acl19disentangle}. Their model is trained from 77,563 manually annotated utterances of disentangled dialogs from online chatting. It is a feedforward neural network with 2 layers, 512-dimensional hidden vectors, and softsign non-linearities. The input of the model is a 77-dimensional vector, where each element is a numerical feature extracted from the original conversation texts, that include time intervals from previous chat utterances posted by the current user, is there a target user in the chat content, do two chat texts contain the same words and so on. 
In Kummerfeld et al.'s study, the model is reported to achieve relatively good performance with 74.9\% precision and 79.7\% recall.

\subsection{Utterance Embedding Layer}
This layer aims to encode not only textual information of utterances but also capture their contextual information. 

\textbf{Utterance Encoding.}
First, for all the utterances in one dialog $D=[u_h, u_{b1}, u_{b2}, ..., u_{bn}]$, we encode it using a pre-trained BERT model \cite{bertbase}, as BERT has been proved to be successful in many natural language processing tasks \cite{bert_classification,bert_ner}. The BERT model is a bidirectional transformer using a combination of \textit{Masked Language Model} and \textit{Next Sentence Prediction}. It is trained from English Wikipedia with nearly 2,500M words.  
The BERT embedding layer outputs $\vec{u}\in\mathbb{R}^d$, which is an 800-dimensional vector for each utterance.

\textbf{Local Window Context.}
Second, we model the contextual information of an utterance through the concept of the \textit{local window}, and use the size of the local window as a hyper-parameter.
Intuitively, the consecutive reply of an issue utterance may be very different from that of non-issue ones. Therefore, we construct a local window context to characterize the dynamic contextual information for extracting desired utterances in a dialog. Specifically, we use a fixed-length local window to integrate context, and define the local window of the utterance $\vec{u_i}$ as $\vec{win_i}$ by joining the $u_i$ with its preceding and following $k$ neighbor utterances. The fixed length is $2k+1$.

\begin{equation}
    \vec{win_i}=\{\vec{u_{i-k}}, ..., \vec{u_{i-1}}, \vec{u_i}, \vec{u_{i+1}}, ..., \vec{u_{i+k}}\}
\end{equation}

When the windows are out of bound, we utilize the \textit{Zero Padding} \cite{InforSeek_User_Intent_Pred} to map the fixed length. In this study, we choose $k=1$ for the local window.

\subsection{Dialog Embedding Layer}

This layer aims to encode utterance features in a multi-faced way to more comprehensively represent the live chat context. To achieve that, we define and extract features from three categories, including textual, heuristic, and contextual features.

\subsubsection{Textual Feature Extractor}
To learn basic textual features for each utterance,
we first represent utterances using TextCNN \cite{kim2014convolutional}. It is a classical method for sentence modeling by using a shallow Convolution Neural Network (CNN) \cite{krizhevsky2012imagenet} to model sentence representation. 
It has an advantage over learning on insufficient labeled data, since it employs a concise network structure and a small number of parameters.
TextCNN uses several convolution kernels to capture local information as the receptive field. Then the global representation is produced with the local information.

Given a kernel $\vec{w}\in\mathbb{R}^h$ with kernel size $h$ and word embedding $\vec{x}=\vec{u_i}$, one convolution feature $\gamma_t$ is generated with $\vec{x}_{t:t+h-1}$:
\begin{equation}
    \gamma_t=\text{ReLU}(\vec{w}\cdot\vec{x}_{t:t+h-1}+b)
\end{equation}
where $b\in\mathbb{R}$ is the bias parameter, and \text{ReLU} is the activate function. We concatenate all the $\gamma_t$ as a feature map:
\begin{equation}
    \vec{\gamma}=[\gamma_1, ..., \gamma_t, ..., \gamma_{n-h+1}]
\end{equation}
where the vector $\vec{\gamma}\in\mathbb{R}^{n-h+1}$. We use \textit{Max-Pooling} strategy to calculate $\hat{\gamma}=\text{max}(\vec{\gamma})$. We set the number of kernels as $m$, and input $u_i$ into three \textit{Convolution-Pooling} layers.
The kernel number of the three layers are 1024, 512, and 256, respectively.
The output of each layer is  $\vec{\Gamma}\in\mathbb{R}^m$, which is a 256-dimensional textual feature vector.

\begin{equation}
    \vec{\Gamma}=[\hat{\gamma}_1, \hat{\gamma}_2, ..., \hat{\gamma}_m]
\end{equation}

\subsubsection{Heuristic Attribute Extractor}

The heuristic attribute extractor aims to augment the dialog embedding results by incorporating high-level semantic attributes from five aspects: \textit{Keyword}, \textit{Structure}, \textit{Sentiment}, \textit{Topic}, and \textit{Role}, 
as elaborated in Table \ref{tab:heurustics}. 

(1) \textit{Keyword}: The occurrences of indicating words or characters about 5W1H, punctuation, etc.

(2) \textit{Structure}: The structural characteristics of utterances in a dialog, such as the number of tokens and positions of the utterances.

(3) \textit{Topic}: {The intuition of this feature is to distinguish off-topic utterances. We first calculate \textit{TF-IDF} \cite{tfidf} for each unique word in the entire chat. Then We extract the top-10 most frequent words and combine them as a 10-dimensional topic vector $\vec{TD_c}$. Similarly, we also extract top-10 most frequent words from dialog head ($\vec{TD_h}$) and the given utterance ($\vec{TD_u}$). Finally, we calculate the \textit{Euclidean Distance} \cite{kusner2015word} between them as the topic deviation:}
\begin{equation}
\begin{aligned}
    TDH=||\vec{TD_c}-\vec{TD_h}||_2 \\  TDU=||\vec{TD_h}-\vec{TD_u}||_2
\end{aligned}
\end{equation}

(4) \textit{Sentiment}: The sentiment information of the given utterance in terms of positive, intermediate, and negative. 

(5) \textit{Role}: The role of the participant who posts the utterance.

By concatenating the above heuristic attributes, this extractor output a 29-dimensional vector.

\subsubsection{Contextual Feature Extractor}
Contextual feature extractor aims to embed the contextual information for each utterance.
{We use \textit{Local Attention} \cite{attn_raw} to represent the context.
The \textit{Local Attention} mechanism mainly focuses on the impact of the neighbor utterances locate in the same window. \textit{Local Attention} can use low time-memory cost to highly represent the semantic context.
An attention function can be described as mapping a query and a set of key-value pairs to an output \cite{self_attn}, which is a triple: $(\vec{h_Q},\vec{h_K},\vec{h_V})$. The function uses the query vector $\vec{h_Q}$ calculated by the given utterance $\vec{u_i}\in\mathbb{R}^d$ to query the attention scores with key vector $\vec{h_K}$.
The key vector $\vec{h_K}$ can be calculated by each utterance $\vec{u_s}\in\mathbb{R}^d$ within the local window ($i-k\leq s\leq i+k$). The attention weight vector is calculated by multiplying value vector $\vec{h_V}$ and sum the attention value.}
Therefore, we define the trainable query matrix $\bm{W}^Q\in\mathbb{R}^{\delta\times d}$, key matrix $\bm{W}^K\in\mathbb{R}^{\delta\times d}$ and value matrix $\bm{W}^V\in\mathbb{R}^{\delta\times d}$,
and calculate the triple $(\vec{h_Q},\vec{h_K},\vec{h_V})$:
\begin{equation}
\vec{h_Q}=\bm{W}^Q\vec{u_i}, \vec{h_K}=\bm{W}^K\vec{u_s}, \vec{h_V}=\bm{W}^V\vec{u_s}
\end{equation}
where the output contextual weight vector is defined as $\vec{c}_i\in\mathbb{R}^\delta$, which can be calculated by the following equations:
\begin{equation}
\begin{aligned}
    & score(\vec{h_Q}, \vec{h_K})=(\vec{h_Q}\cdot \vec{h_K})\exp{[-\frac{(s-i)^2}{2k^2}]} \\
    & a_s=softmax(\vec{h_Q}, \vec{h_K})=\frac{score(\vec{h_Q}, \vec{h_K})}{\sum_s score(\vec{h_Q}, \vec{h_K})} \\
    & \vec{c_i}= \sum_{i-k\leq s\leq i+k}(a_s\vec{h_V})/\sqrt{d}
\end{aligned}
\end{equation}

{Equation (9) shows the three processes of \textit{Local-Attention} calculation: (1) Output attention vector with dot production as $score(\vec{h_Q}, \vec{h_K})$ by multiplying \textit{Gaussian Distance} between $s$ and $i$; (2) Use \textit{Softmax} to normalize the score vector; and (3) Apply the normalized score vector to calculate the local attention.}
We set the parameter $d=800$, $\delta=128$, and obtain a 128-dimensional context vector.

Finally, we concatenate the vectors that output by the three extractors into a 413-dimensional feature vector $\vec{u'}$.

\subsection{Output Layer}
We input the feature vector $\vec{u'}$ into two \textit{Full-Connected Layers} (FC), and use two \textit{Softmax} functions to calculate the probability of issue-description utterance and the corresponding solution utterances, respectively. 
\begin{equation}
\begin{aligned}
    & P(I|u_h)=softmax(FC_1(\vec{{u_h}'})) \\
    & P(S|u_{bi})=softmax(FC_2(\vec{{u_i}'}))
\end{aligned}
\end{equation}
where $P(I|u_h)$ is the predicted probability of issue-description utterance, and $P(S|u_{bi})$ ($u_{bi}\in\{u_{b1}, ..., u_{bn}\}$) is the predicted probability of solution utterance. The \textit{Cross-Entropy Loss} are applied with the two tasks when measuring the difference between truth and prediction. The two loss functions are defined as $Loss_I$ and $Loss_S$:
\begin{equation}
\begin{aligned}
    & Loss_I=-y_h\cdot\log P(I|u_h) \\
    & Loss_S=-y_{i}\cdot\log P(S|u_{bi})
\end{aligned}
\end{equation}
where $y_h$ and $y_{i}$ indicate the {ground-truth} labels of utterances. The issue model and solutions model are separately trained until convergence. 

\subsection{Application}
When fully trained, for a given chat log, {\tool} automates the three sub-tasks formulated in Section \ref{sec:background}:  
First, it performs dialog disentanglement; Second, for each disentangled dialog, it uses the issue model to predicts whether the dialog head is issue description; and Finally, if the issue model predicts positive, then it uses the trained solution model to predict which utterances can be selected into the solution. As a final output, it combines the predicted utterances as the corresponding solution to the issue.  
\section{Experimental Design}
To evaluate the proposed {\tool} approach, our evaluation specifically addresses three research questions:

\textbf{RQ1}: What is the performance of {\tool} in detecting issue dialogs from live chat data? 

\textbf{RQ2}:  What is the performance of {\tool} in extracting solutions for a given issue? 

\textbf{RQ3}: How does each individual component in {\tool} contribute to the overall performance? 

\vspace{-1ex}
\subsection{Data Preparation}

\subsubsection{Studied Communities}

Many OSS communities utilize Gitter \cite{gitter.im} or Slack \cite{slack.com} as their live communication means. 
Considering the popular, open, and free access nature, we select studied communities from {Gitter}\footnote{In Slack, communities are controlled by the team administrators, whereas in Gitter, access to the chat data is public.}.

To identify studied communities, we select the Top-1 most participated communities from eight active domains, covering front end framework, mobile, data science, DevOps, blockchain platform, collaboration, web app, and programming language. 
Then, we collect the daily chat utterances from these communities. Gitter provides REST API \cite{gitter.rest} to get data about chatting rooms and post utterances. In this study, we use the REST API to acquire the chat utterances of the eight selected communities, and the retrieved dataset contains all utterances as of ``2020-12''.

\subsubsection{Bootstrap Sampling}
After dialog disentanglement, the number of separate chat dialogs is large. Limited by the human resource of labeling, we randomly sample 100 dialogs from each community. Then we excluded unreadable dialogs: 1) Dialogs that are written in non-English languages; 2) Dialogs that contain too much code or stack traces; 3) Low-quality dialogs such as dialogs with many typos and grammatical errors. 4) Dialogs that involve channel robots. 
However, the dataset is imbalanced in that the non-issue dialogs are much more than issue dialogs, as shown in Table \ref{tab:dataset}. Therefore, we apply an arbitrary bootstrap sampling strategy \cite{bootstrap} for data balancing by randomly sampling issue dialogs with replacement until the number of issue dialogs and non-issue dialogs is balanced. 

\subsubsection{Ground-truth Labeling}
For each sampled dialog, we first manually label whether its head discussed a certain issue. Then, for each issue-dialog, we label the utterances that should be included in the solution. 
The labeled results are used as the \textbf{ground-truth} dataset for performance evaluation. 
To guarantee the correctness of the labeling results, we built an inspection team, which consisted of four Ph.D. candidates. All of them are fluent English speakers, and have done either intensive research work with software development or have been actively contributing to open-source projects. We divided the team into two groups. The labeled results from the Ph.D. candidates were reviewed by others. When a labeled result received different opinions, we hosted a discussion with all team members to decide through voting. The average Cohen’s Kappa about issue-dialog is 0.85, and the average Cohen’s Kappa about solution-utterance is 0.83.

\begin{table}[tp]
\caption{The statistics of the eight Gitter communities. 
}
 \resizebox{\columnwidth}{16mm}{
\begin{tabular}{|c|c|c|c|c|c|c|c|c|}
\hline
\multirow{2}{*}{\textbf{Id}}&\multirow{2}{*}{\textbf{Project}} & \multicolumn{3}{c|}{\textbf{Entire Population}} & \multicolumn{4}{c|}{\textbf{Sample Population}} \\
\cline{3-9}
& & \textbf{Par.} & \textbf{Dial.} & \textbf{Utter.}& \textbf{Issue} & \textbf{Utter.} & \textbf{Non-Issue} & \textbf{Utter.} \\ \hline
$P_1$& Angular  & 22,467  & 79,619  & 695,183  & 17 & 170 & 80 & 577\\ \hline
$P_2$& Appium  & 3,979    & 4,906  & 29,039   & 27 & 461 & 60 & 404\\ \hline
$P_3$& Dl4j & 8,310  & 27,256 & 252,846  & 38 & 665 & 55 & 1,123\\ \hline
$P_4$& Docker & 8,810   & 3,954   & 22,367 & 20 & 151 & 74 & 352 \\ \hline
$P_5$& Ethereum & 16,154 & 17,298  & 91,028  & 17 & 115 & 83 & 478 \\ \hline
$P_6$& Gitter  & 9,260  & 7,452 & 34,147  & 22  & 315 & 64 & 575 \\ \hline
$P_7$& Nodejs  & 18,118 & 13,981 & 81,771 & 14 & 121 & 81 & 359 \\ \hline
$P_8$& Typescript  & 8,318  & 18,812 & 196,513  & 16 & 361 & 82 & 1,076\\ \hline

\multicolumn{2}{|c|}{\textit{Total}} & \textit{95,416}        & \textit{173,278}  & \textit{1,402,894}    & \textit{171}  & \textit{2,359} & \textit{579} & \textit{4,944}\\ \hline
\end{tabular}}

\vspace{-0.5cm}
\label{tab:dataset}
\end{table}

In total, we collected 173,278 dialogs from eight open-source communities, and spent 720 person-hours on annotating {\numdataset} dialogs including 171 issue-solution pairs. Table \ref{tab:dataset} presents the detail of our dataset. It shows the number of participants (Par.), dialog (Dial.), and utterance (Utter.)  for the entire population, as well as the number of issue and non-issue dialogs with the corresponding utterances for the sample population.
Moreover, to contribute to the eight communities, we apply {\tool} on the 173,278 dialogs. We extract and publish {\numPairExp} issue-solution pairs on our website.

\subsection{Baselines}

The first two RQs require the comparison of {\tool} with state-of-the-art baselines. Due to the slightly different focuses between RQ1 and RQ2, we employ three common baselines applicable for both, as well as three additional baselines for each RQ. This leads to a total of six baselines for each RQ.

{\textbf{Common Baselines applicable for RQ1 and RQ2.}} The three commonly used machine-learning-based baselines 
are utilized to comprehensively examine the classification performance, i.e., {\textbf{Naive Bayesian (NB)}} \cite{mccallum1998comparison}, {\textbf{Random Forest (RF)}} \cite{liaw2002classification}, and {\textbf{Gradient Boosting Decision Tree (GBDT)}} \cite{ke2017lightgbm}.

{\textbf{Additional Baselines for detecting issues (RQ1).}}
\textbf{Casper \cite{DBLP:conf/icse/0002S20}}
is a method for extracting and synthesizing user-reported mini-stories regarding app problems from reviews. We use utterances as the extracted events, and treat its second step, i.e., classify problems, as a baseline. 
We use the implementation provided by the original paper \cite{Casper}.
\textbf{CNC\_PD \cite{Huang2018Automating}}
is the state-of-the-art learning technique to classify sentences in comments taken from online issue reports. They proposed a CNN \cite{1998Neural}-based approach to classify sentences into seven categories of intentions: Feature Request, Solution Proposal, Problem Discovery, etc. 
we treat the CNN classifier that predicts utterances as the \textit{Problem Discovery} category as a baseline for detecting issues.
\textbf{DECA\_PD} \cite{DBLP:conf/kbse/SorboPVPCG15}
is the state-of-the-art rule-based technique 
for analyzing development email content. It is used to classify the sentences of emails
into problem discovery, solution proposal, information giving, etc., by using linguistic rules. We use the six linguistic rules \cite{deca_web} for identifying the ``problem discovery" dialog-head as our baseline.

{\textbf{Additional baselines for extracting solutions (RQ2).}}
\textbf{UIT} \cite{InforSeek_User_Intent} is a context-representative classifier that uses Glove \cite{glove} to embed words, and uses TextCNN to embed the utterance. The UIT classifies utterance into 12 categories. Specifically, we choose the ``potential answer" classifier as a solution-extraction baseline.
\textbf{CNC\_SP} is the \textit{Solution Proposal} classifier in \cite{Huang2018Automating}.
\textbf{DECA\_SP} is the set of 51 linguistic rules for identifying ``solution proposal" sentences in \cite{DBLP:conf/kbse/SorboPVPCG15}.

\subsection{Evaluation Metrics}
\begin{table*}[htbp]
\caption{Baseline comparison across the eight communities (\%). }
\centering
 \resizebox{\textwidth}{22mm}{
\begin{tabular}{|c|c|c|c|c|c|c|c|c|c|c|c|c|c|c|c|c|c|c|c|c|c|c|c|c|c|c|c|c|}
\hline
\multirow{2}{*}{\textbf{Task}} & \multirow{2}{*}{\textbf{Methods}} & \multicolumn{3}{c|}{\textbf{Angular}} & \multicolumn{3}{c|}{\textbf{Appium}} & \multicolumn{3}{c|}{\textbf{Docker}} & \multicolumn{3}{c|}{\textbf{DL4j}} & \multicolumn{3}{c|}{\textbf{Ethereum}} & \multicolumn{3}{c|}{\textbf{Gitter}} & \multicolumn{3}{c|}{\textbf{Nodejs}} & \multicolumn{3}{c|}{\textbf{Typescript}} &
\multicolumn{3}{c|}{\textbf{Average}}\\
\cline{3-29}
& & P & R & F1 & P & R & F1 & P & R & F1 & P & R & F1 & P & R & F1 & P & R & F1 & P & R & F1 & P & R & F1 & P & R & F1  \\
\hline
\multirow{7}{*}{Issue} & {\tool} & \cellcolor{CadetBlue2}76 & \cellcolor{CadetBlue2}77 & \cellcolor{CadetBlue2}76 & \cellcolor{CadetBlue2}75 & 68 & \cellcolor{CadetBlue2}71 & \cellcolor{CadetBlue2}84 & 74 & \cellcolor{CadetBlue2}79 & 77 & \cellcolor{CadetBlue2}68 & \cellcolor{CadetBlue2}72 & 82 & 73 & \cellcolor{CadetBlue2}77 & \cellcolor{CadetBlue2}80 & \cellcolor{CadetBlue2}69 & \cellcolor{CadetBlue2}74 & \cellcolor{CadetBlue2}79 & 70 & \cellcolor{CadetBlue2}74 & \cellcolor{CadetBlue2}86 & 78 & \cellcolor{CadetBlue2}82 & \cellcolor{Cyan3}80 & \cellcolor{Cyan3}72 & \cellcolor{Cyan3}76\\
\cline{2-29} & NB & 36 & 40 & 38 & 41 & 30 & 35 & 47 & 36 & 41 & 70 & 56 & 62 & 08 & 25 & 13 & 22 & 42 & 29 & 30 & 50 & 37 & 15 & 40 & 22 & 34 & 40 & 36 \\
\cline{2-29} & RF & 56 & 25 & 34 & 69 & 30 & 42 & 75 & 23 & 35 & \cellcolor{CadetBlue2}84 & 44 & 58 & \cellcolor{CadetBlue2}100 & 17 & 29 & 50 & 25 & 33 & 33 & 13 & 18 & 23 & 30 & 26 & 61 & 26 & 36 \\
\cline{2-29} & GBDT & 27 & 75 & 40 & 40 & \cellcolor{CadetBlue2}70 & 51 & 50 & \cellcolor{CadetBlue2}79 & 61 & 73 & 44 & 55 & 21 & \cellcolor{CadetBlue2}76 & 33 & 19 & 67 & 29 & 30 & \cellcolor{CadetBlue2}88 & 44 & 18 & \cellcolor{CadetBlue2}90 & 30 & 35 & 65 & 46 \\
\cline{2-29} & Casper & 39 & 35 & 37 & 08 & 03 & 05 & 59 & 26 & 36 & 46 & 40 & 43 & 19 & 42 & 26 & 14 & 17 & 15 & 05 & 06 & 06 & 15 & 40 & 22 & 26 & 26 & 26 \\
\cline{2-29} & CNC\_PD & 20 & 55 & 29 & 23 & 50 & 32 & 23 & 36 & 28 & 12 & 32 & 17 & 24 & 42 & 30 & 12 & 42 & 19 & 10 & 50 & 17 & 05 & 40 & 10 & 16 & 43 & 24\\
\cline{2-29} & DECA\_PD & 33 & 50 & 40 & 28 & 37 & 31 & 33 & 36 & 34 & 64 & 28 & 39 & 42 & 42 & 42 & 44 & 67 & 53 & 32 & 50 & 39 & 04 & 10 & 06 & 35 & 40 & 37\\
\hline
\multirow{7}{*}{Solution} & {\tool} & \cellcolor{CadetBlue2}61 & 58 & \cellcolor{CadetBlue2}69 & \cellcolor{CadetBlue2}68 & 57 & \cellcolor{CadetBlue2}62 & \cellcolor{CadetBlue2}71 & 60 & \cellcolor{CadetBlue2}65 & \cellcolor{CadetBlue2}66 & \cellcolor{CadetBlue2}62 & \cellcolor{CadetBlue2}58 & \cellcolor{CadetBlue2}73 & 63 & \cellcolor{CadetBlue2}68 & \cellcolor{CadetBlue2}62 & 56 & \cellcolor{CadetBlue2}59 & \cellcolor{CadetBlue2}68 & 58 & \cellcolor{CadetBlue2}63 & \cellcolor{CadetBlue2}72 & 67 & \cellcolor{CadetBlue2}69 & \cellcolor{Cyan3}68 & 59 & \cellcolor{Cyan3}63\\
\cline{2-29} & NB & 21 & 58 & 30 & 24 & 48 & 32 & 31 & 59 & 40 & 58 & 49 & 53 & 33 & \cellcolor{CadetBlue2}80 & 47 & 10 & \cellcolor{CadetBlue2}67 & 17 & 37 & 55 & 44 & 08 & 09 & 09 & 28 & 53 & 37 \\
\cline{2-29} & RF & 26 & 83 & 39 & 27 & 52 & 36 & 31 & 56 & 40 & 80 & 13 & 22 & 15 & 40 & 22 & 10 & 33 & 15 & 50 & 65 & 57 & 54 & 64 & 58 & 37 & 51 & 43 \\
\cline{2-29} & GBDT & 26 & \cellcolor{CadetBlue2}92 & 40 & 26 & \cellcolor{CadetBlue2}74 & 38 & 22 & \cellcolor{CadetBlue2}62 & 32 & 71 & 16 & 26 & 11 & 40 & 17 & 10 & 67 & 17 & 42 & \cellcolor{CadetBlue2}75 & 54 & 42 & \cellcolor{CadetBlue2}73 & 53 & 31 & \cellcolor{Cyan3}62 & 41 \\
\cline{2-29} & UIT & 29 & 17 & 21 & 21 & 13 & 17 & 30 & 12 & 17 & 24 & 07 & 12 & 37 & 16 & 22 & 27 & 11 & 17 & 19 & 15 & 18 & 31 & 18 & 23 & 27 & 14 & 18 \\
\cline{2-29} & CNC\_SP & 18 & 17 & 17 & 26 & 28 & 27 & 21 & 31 & 25 & 46 & 20 & 28 & 30 & 14 & 19 & 46 & 54 & 50 & 41 & 28 & 33 & 38 & 25 & 30 & 33 & 27 & 30 \\
\cline{2-29} & DECA\_SP & 00 & 00 & 00 & 00 & 00 & 00 & 20 & 09 & 12 & 25 & 02 & 24 & 00 & 00 & 00 & 00 & 00 & 00 & 57 & 20 & 30 & 33 & 09 & 14 & 17 & 05 & 08 \\
\hline
\end{tabular}}
\label{tab:baseline}
\end{table*}

We use three commonly-used metrics to evaluate the performance, i.e., \textit{Precision, Recall, F1}.
(1) \textit{Precision}, which refers to the ratio of the number of correct predictions to the total number of predictions; 
(2) \textit{Recall}, which refers to the ratio of the number of correct predictions to the total number of samples in the golden test set; and (3) \textit{F1}, which is the harmonic mean of precision and recall. When comparing the performances, we care more about F1 since it is balanced for evaluation.
Note that, since the number of utterances may largely vary across different dialogs, we calculate the performance of solution extraction in the scope of the community.

\subsection{Experiment Settings}

For all experiments, we apply \textit{Cross-Project Evaluation} on our dataset to perform the training process. We iteratively select one project as testset, and the remaining seven projects for training.
{The experiment environment is a Windows 10 desktop computer, NVIDIA GeForce RTX 2060 GPU, intel core i7, and 32GB RAM.}

To answer RQ1, we first train the issue dataset with 8 \textit{batch\_size}. Each of the convolution and dense layers use 0.6 \textit{dropout} to avoid overfitting. The optimizer chooses \textit{Adam}=0.001 and $\beta_1$=0.9. We train {\tool} for 100 epochs with 5 patience \textit{Early-Stopping}. We set the threshold of predicted possibility within 0.2-0.8, and choose \textit{threshold}=0.5 as positive-negative boundary {which can reach the best performance after tuning}. NB/GDBT/RF baselines choose the default parameter settings for training; Casper chooses \textit{SVM.SVC} as default function, with \textit{rbf} as kernel, 3 as degree, and 200 as \textit{cache\_size}; CNC\_PD selects 64 as \textit{batch\_size}, 192-dimensional word embedding, four different filter sizes of $[2,3,4,5]$ with 128 filters, 50 training epochs 
and \textit{dropout}=0.5.
These baseline parameters are determined by a greedy strategy, and can achieve the best performance after tuning.

For RQ2, we use the same parameters with RQ1 for {\tool}. We only change the predicting threshold from 0.5 to 0.4 since 0.4 can reach the best performance on solution extraction after tuning. For baselines, we choose default parameter settings for NB/GDBT/RF training. For UIT, we select 32 as \textit{batch\_size} and 0.6 as \textit{dropout}.

For RQ3, 
we compare {\tool} with its three variants: 1) \textbf{{\tool}-CNN}, which removes the textual feature extractor from {\tool}, 2) \textbf{{\tool}-Heu}, which removes the hueristic attribute extractor from {\tool}, and 3) \textbf{{\tool}-LocalAttn}, which removes the contextual feature extractor from {\tool}. 
Three variants use the same parameters when training.


\section{Results}
\subsection{Performance in Detecting Issues}

\begin{figure*}[t]
\centering
\includegraphics[width=\textwidth]{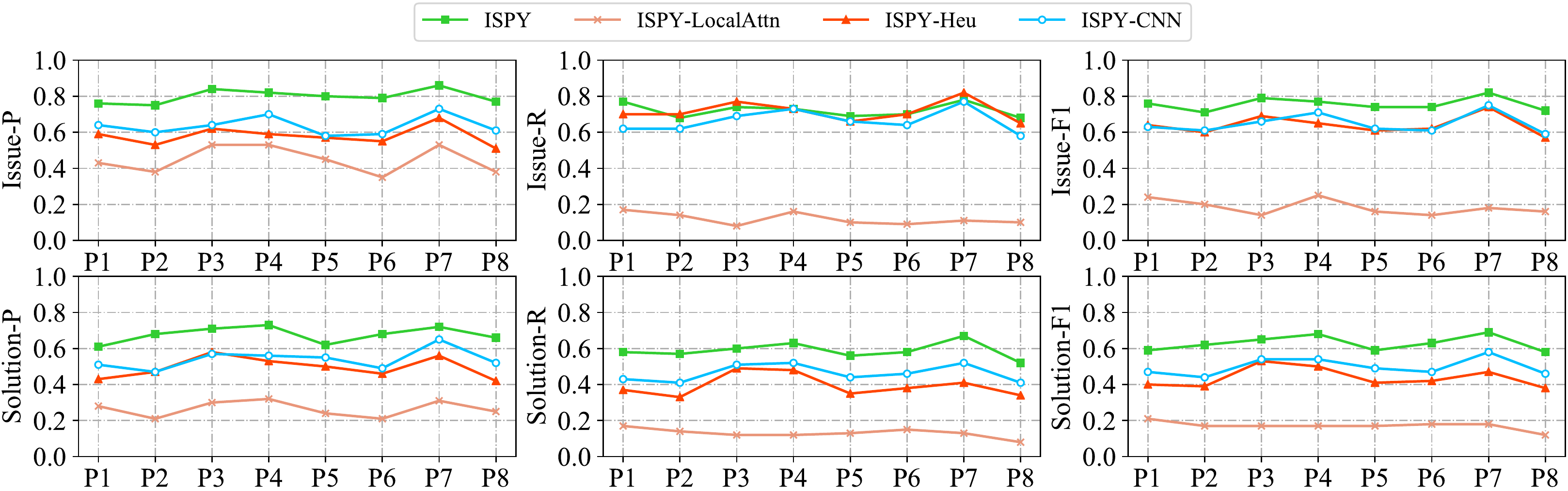}
\caption{The component analysis.}
\label{fig:ablation}
\vspace{-0.5cm}
\end{figure*}

The upper half of Table \ref{tab:baseline} demonstrates the comparison results between the performance of {\tool} and those of the six baselines
across data from eight OSS communities, for \textbf{issue detection} tasks.
The columns correspond to Precision, Recall, and F1 score. 
The highlighted cells indicate the best performances from each column. 
Then, we conduct the normality test and T-test between every two methods.
Overall, the data follow a normal distribution ($p=0.32$)\footnote{Significant test: $p<0.05$}, and {\tool} significantly ($p=10^{-20}$)
outperforms the six baselines in terms of the average Precision, Recall, and F1 score. 
Specifically, when comparing with the best Precision-performer among the six baselines, i.e., RF, {\tool} can improve 
its average precision by 19\%. Similarly, {\tool} improves the best Recall-performer, i.e., GBDT, by 7\% for average recall, and improves the best F1-performer, i.e., GBDT, by 30\% for average F1 score. At the individual project level, {\tool} can achieve the best performances on most of the eight communities. These results indicate that {\tool} can more accurately detect whether a dialog is discussing an issue, than all comparison baselines. 

We believe that the performance advantage of {\tool} is mainly attributed to the rich representativeness of its internal construction, from two perspectives: 
(1) {\tool} can accurately capture the semantic relationship between the issue description and its first reply, by using the local window and local attention mechanism. This enables it to learn more comprehensive contextual knowledge, e.g., what kind of first issue-replies represents a dialog head containing issue descriptions. Therefore, it contributes to more accurate classification. 
(2) {\tool} augments the textual vectors with high-level semantic information by employing {\attribute} heuristic attributes. 
For example, three out of the {\attribute} attributes are characterizing sentiment attributes. This design is based on the observation that issue descriptions are likely to contain negative tones such as ``fail'', ``error'', and ``annoy''. 
By calculating three types of polarity sentiment scores as shown in Table \ref{tab:heurustics},
the sentiment attributes can be fit into the deep learning network to help with issue-description detection.

\textbf{Answering RQ1:} {\tool} outperforms the six baselines in detecting issue dialogs across most of the studied projects, and the average Precision, Recall, and F1 are 80\%, 72\%, and 76\%, respectively, improving the best F1-baseline GBDT by 30\% on average F1 score.

\subsection{Performance in Extracting Solutions}

\begin{figure}[t]
\centering
\includegraphics[width=\columnwidth,height=10cm]{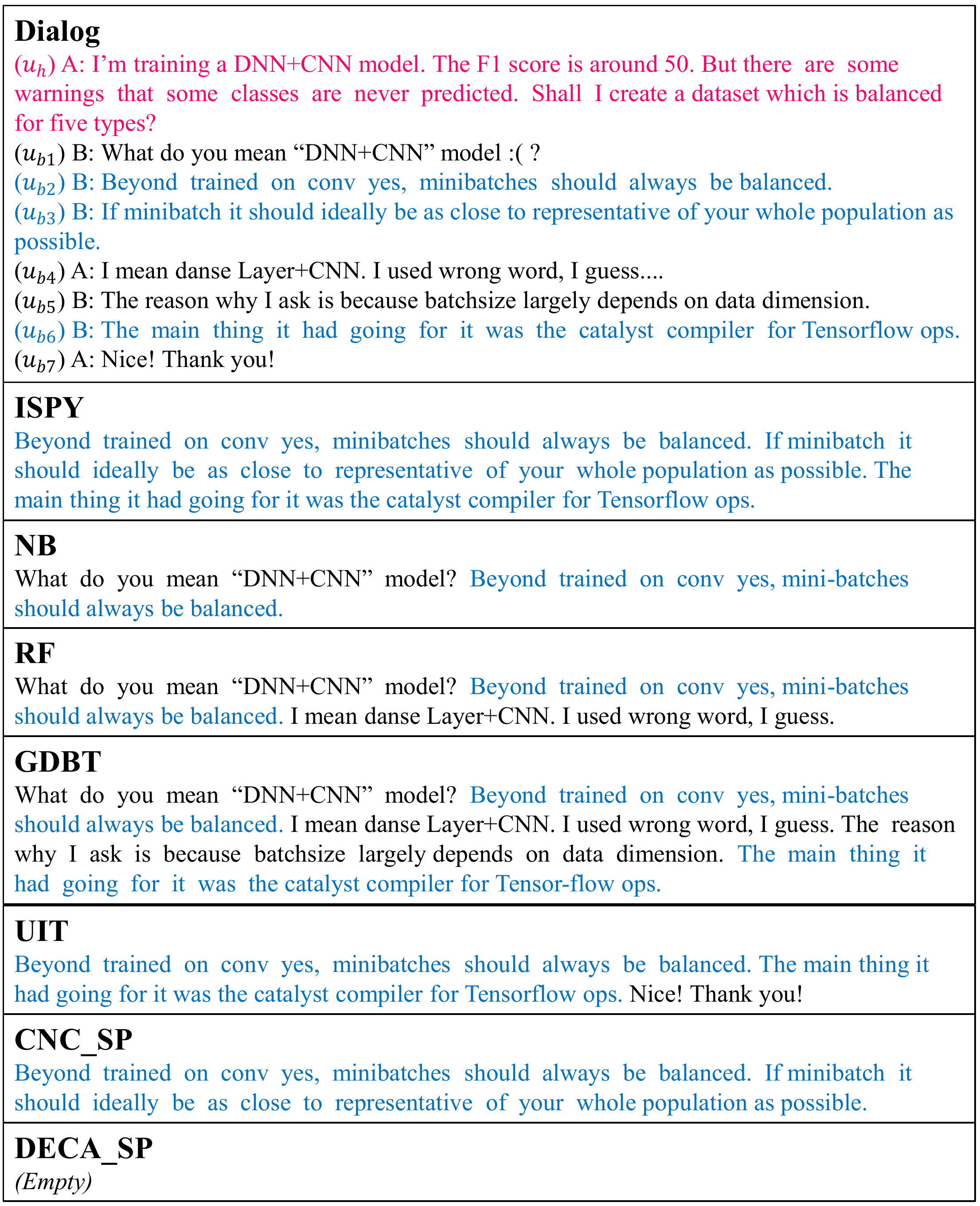}
\caption{Test example.}
\label{tab:example_baseline}
\vspace{-0.8cm}
\end{figure}

Similarly, the bottom part of Table \ref{tab:baseline} summarizes the comparison results between the performance of {\tool} and those of the six baselines across data from eight OSS communities, for \textbf{solution extraction} task. 
We can see that, {\tool} can achieve the highest performance in most of the columns. It significantly ($p=10^{-5}$) outperforms the six baselines. On average, although {\tool} are slightly below GBDT by 3\% of Recall, it reaches the highest F1 score (63\%), improving the best baseline RF by 20\%. It also reaches the highest precision (68\%), significantly higher than other baselines (i.e., ranging from 17\% to 37\%). These results imply that {\tool} can effectively extract utterances as the corresponding solutions from development dialogs:
(1) Our approach is sensitive to identifying solutions, including consecutive utterances, by employing a local attention mechanism. In live chats, some solutions include consecutive utterances from the same participants. For example, the utterance $u_{b2}$ and $u_{b3}$ in Figure \ref{tab:example_baseline} are both selected as solution utterances. Our approach can learn that knowledge by adjusting the weight of $u_{b3}$ to be higher, to increase its probability according to the local attention learning. While baselines (e.g., NB, RF, GDBT, UIT, DECA\_SP) separately learn the textual characteristics of each utterance, thus are more prone to predict $u_{b3}$ as a non-solution utterance.   
(2) Our approach can screen negative feedbacks in consecutive utterances and reject ineffective solutions 
based on the heuristic attributes 
(i.e., disapproval keywords and negative sentiment attributes)
and local attention mechanism. In live chats, some utterances are indeed solutions but are proved ineffective (e.g., those have follow-up utterances like ``It doesn't work'') by the dialog initiator later. In such cases, {\tool} can detect whether its follow-up utterances contain negative feedback based on the heuristic attributes and local attention learning, while other methods cannot.

We also notice that, the DECA\_SP baseline can hardly extract the solution utterances correctly. By investigating their 51 linguistic rules and our test dataset, we consider that it comes from two reasons. First, the DECA\_SP rules are designed for extracting solution proposals from email contents, which have different expressing styles from live chats. Second, the DECA\_SP rules are kind of strict for live chats. For example, the rule ``[something] can be fixed by [something]'' cannot deal with its similar variants such as ``[something] could/should be fixed by [something]''.


\textbf{Answering RQ2:} {\tool} outperforms the six baselines in extracting solution utterances in terms of Precision and F1. The average Precision, Recall, and F1 are 68\%, 59\%, and 63\%, respectively, improving the best F1-baseline RF by 20\% on average F1 score. 

\subsection{Effects of Main Components}

Fig. \ref{fig:ablation} presents the performances of {\tool} and its three variants respectively. We can see that, the F1 performances of {\tool} are higher than all three variants in both issue-dialog detection and solution extraction tasks.
When compared with {\tool} and {\tool}-LocalAttn, removing the LocalAttn component will lead to a dramatic decrease of the precision (-35\%), recall (-60\%) , and F1 score (-57\%) for the solution-utterance extraction task, as well as a decrease of the precision (-41\%), recall (-46\%) , and F1 score (-46\%) for the solution extraction task. This indicates that the local attention mechanism is an essential component to contribute to {\tool}'s high performance in both detecting issues and extracting solution utterances.
The top three charts in Fig. \ref{fig:ablation} compare the precision, recall, and F1-score of {\tool} and its three variants, for the issue-detection task. Compared to {\tool}-Heu and {\tool}-CNN, {\tool} has moderately better precision and F1, and the recalls of all the three remain very close. 
It is because that, the contextual information is quite effective in retrieving all the positive-truth instances back for this task, while the other two components mainly contribute to filter the negative-truth instances out, thus can further improve precision. 
The bottom three charts in Fig. \ref{fig:ablation} compare the precision, recall, and F1-score of {\tool} and its three variants, for the solution-extraction task. Compared to {\tool}-Heu and {\tool}-CNN, {\tool} has moderately better precision, recall, and F1.

\textbf{Answering RQ3:} The textual feature extractor, heuristic attribute extractor, and content feature extractor adopted by {\tool} are helpful for extracting issue-solution pairs, while the contextual feature extractor provides a more significant contribution to the effectiveness of {\tool} than others.

\section{An Application Study of {\tool}}

Experiments in Section V have shown the performances of our approach. In this section, we conduct an application study to
further demonstrate the usefulness of our approach. 

\textbf{Procedure}. We apply {\tool} on live chat data from three new communities: Materialize, Springboot, and WebPack (note that these are different from our studied communities). According to the issue-solution pairs extracted from the three communities, we manually inspect their recently unanswered questions on Stack Overflow, and provide potential solutions correspondingly. 
First, we crawl the recent (January 2018 to April 2021) live chats of three new communities. Second, we 
apply {\tool} to disentangle the live chats into about 21K dialogs, and generate a dataset with over 9K issue-solution pairs. Because all live chats are historical data, we cannot directly evaluate the usefulness of {\tool} with the original developers who inquired about the issue. As an alternative, we investigate the usefulness of {\tool} by sharing the discovered solutions to developers facing similar issue on Stack Overflow. Specifically, we employ four Ph.D. students to inspect the recent, unanswered questions in these three communities on Stack Overflow. When finding unanswered issues that have been discussed in live chats, we post the corresponding solutions as potential answers. 

\begin{table}[bp]
\caption{The statistics of {\tool}'s contribution on answering questions for Stack Overflow. 
}
\resizebox{\columnwidth}{!}{
\begin{tabular}{|c|c|m{4.8cm}<{\centering}|}
\hline
\textbf{Contribution Type} & \textbf{\#} & \textbf{QID (\#ans)} \\ \hline
Accepted Answer & 6 & \href{https://stackoverflow.com/questions/67201337/i-get-a-error-java-lang-noclassdeffounderror}{\textcolor{Turquoise3}{67201337(1)}}, \href{https://stackoverflow.com/questions/67086241/does-anyone-here-know-if-you-can-use-the-new-asset-modules-inline-like-you-could/67119576}{\textcolor{Turquoise3}{67086241(1)}}, \href{https://stackoverflow.com/questions/67101118/how-to-realize-bundle-splitting-that-i-want-do-split-one-bundle-file-into-two-bu}{\textcolor{Turquoise3}{67101118(1)}}, \href{https://stackoverflow.com/questions/67101118/how-to-realize-bundle-splitting-that-i-want-do-split-one-bundle-file-into-two-bu}{\textcolor{Turquoise3}{67047528(1)}}, \href{https://stackoverflow.com/questions/66899397/whether-the-usage-of-nativewebrequest-is-thread-safe-or-not}{\textcolor{Turquoise3}{66899397(1)}}, \href{https://stackoverflow.com/questions/67150784/get-org-springframework-boot-info-buildproperties/67151141}{67150784(2)}\\ \hline
Potential Answer & 19 & \href{https://stackoverflow.com/questions/61529948/how-to-resolve-this-loader-issue-in-react/66868053}{\textcolor{Turquoise3}{61529948(1)}}, \href{https://stackoverflow.com/questions/65619783/transpiling-to-es5-using-webpack-4-and-babel/66868545}{\textcolor{Turquoise3}{65619783(1)}}, \href{https://stackoverflow.com/questions/63620416/webpack-imports-all-the-unused-exports-into-the-bundle/66980692}{\textcolor{Turquoise3}{63620416(1)}}, \href{https://stackoverflow.com/questions/66383230/error-couldnt-find-a-style-target-it-started-to-appear-after-ssr-with-react-i}{\textcolor{Turquoise3}{66383230(1)}}, \href{https://stackoverflow.com/questions/57543742/org-springframework-boot-builder-springapplicationbuilder-initljava-lang-obj/67070851}{\textcolor{Turquoise3}{57543742(1)}}, \href{https://stackoverflow.com/questions/49303635/how-can-i-force-autoconfigurejsontesters-to-use-hal-format-from-spring-hateoas/67111285}{\textcolor{Turquoise3}{49303635(1)}}, \href{https://stackoverflow.com/questions/51042100/failed-to-instantiate-sessionfactory-from-org-springframework-orm-hibernate5-loc/67119179}{\textcolor{Turquoise3}{51042100(1)}}, \href{https://stackoverflow.com/questions/39653049/how-to-combine-materializecss-slider-and-parallax/67177562}{\textcolor{Turquoise3}{39653049(1)}}, \href{https://stackoverflow.com/questions/65877801/materializecss-select-method-doesnt-work-on-ios/67179260}{\textcolor{Turquoise3}{65877801(1)}}, \href{https://stackoverflow.com/questions/64455428/added-non-passive-event-listener-to-a-scroll-blocking-touchmove-event-conside/67194566}{\textcolor{Turquoise3}{64455428(1)}}, \href{https://stackoverflow.com/questions/59282213/react-warning-non-passive-event-listener-to-a-scroll-blocking-touchstart}{59282213(2)}, \href{https://stackoverflow.com/questions/59073683/can-i-use-es6-modules-import-export-to-write-webpack-config-files/66980357}{59073683(2)}, \href{https://stackoverflow.com/questions/63190114/call-a-python-script-from-react-with-next-routing-and-a-node-js-server/66980504}{63190114(2)}, \href{https://stackoverflow.com/questions/50233325/bean-definition-is-overriden-by-autoconfiguration/67111888}{50233325(2)}, \href{https://stackoverflow.com/questions/56158365/register-caffeine-cache-in-spring-actuator-cachemanager/67188973}{56158365(2)}, \href{https://stackoverflow.com/questions/60062706/materializecss-date-picker-is-not-working/67179000}{60062706(3)}, \href{https://stackoverflow.com/questions/42130187/how-to-add-search-for-select-in-materialize-css-framework/67195549}{42130187(3)}, \href{https://stackoverflow.com/questions/38848742/material-select-blinking-on-ios/67179533}{38848742(4)}, \href{https://stackoverflow.com/questions/47974757/webclient-vs-resttemplate/67056700}{47974757(5)}\\ \hline
Comment & 1 & \href{https://stackoverflow.com/questions/66378139/spring-boot-2-4-3-actuator-startup-endpoint-not-found/66378380}{66378139(2)}\\ \hline

\textbf{Total} & \multicolumn{2}{|c|}{\textbf{26}}\\ \hline
\end{tabular}}
\label{tab:so}
\end{table}

\begin{figure}[b]
\centering
\includegraphics[width=\columnwidth, height=5.5cm]{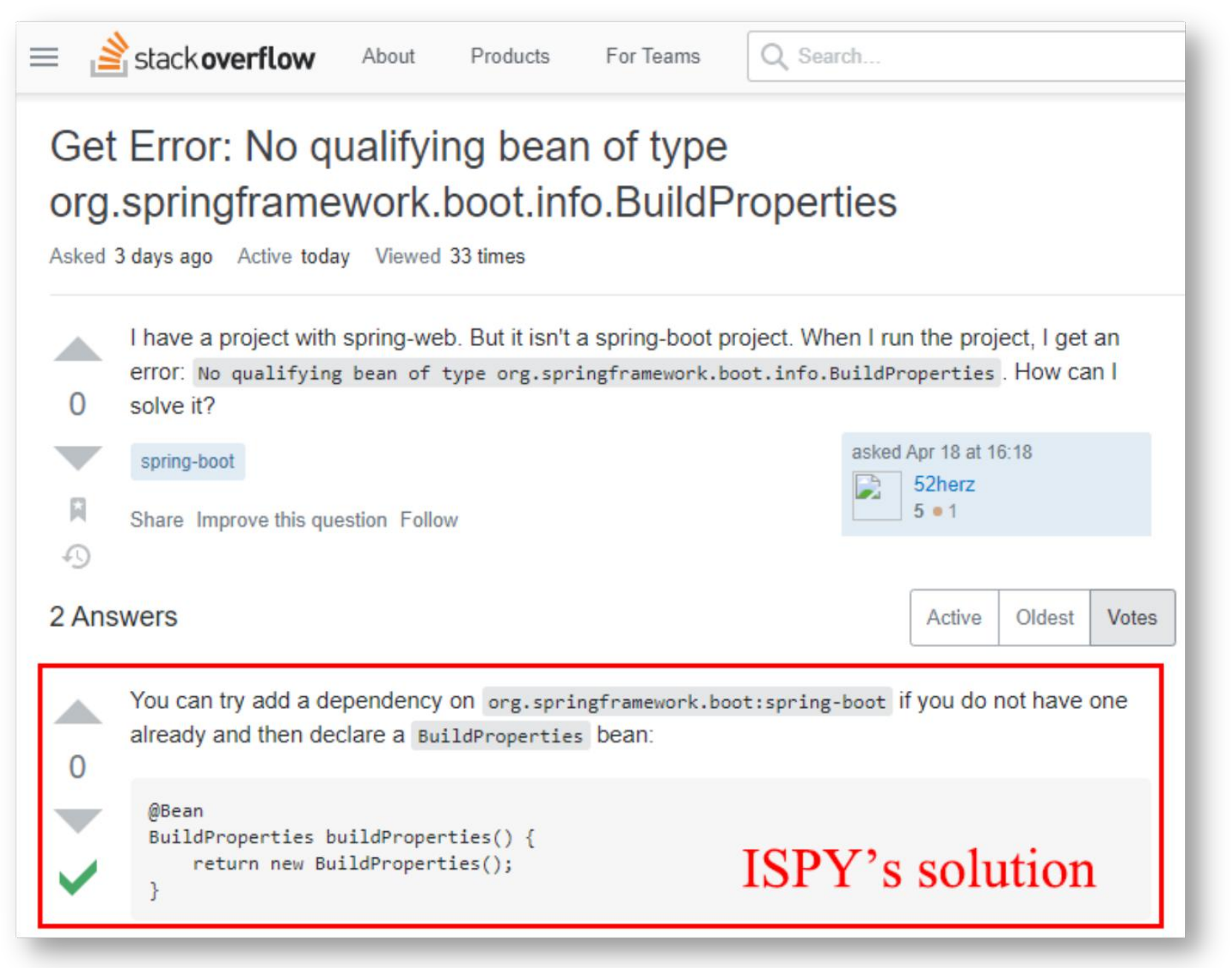}
\caption{
An example of ``BuildProperties" issue resolved by {\tool}'s solution on Stack Overflow}
\label{fig:rq4}
\end{figure}

\textbf{Results.}
Table \ref{tab:so} summarizes the 
results from this application study (More details can be found in our website).
The ``Contribution Type" column shows how {\tool}'s solutions contribute to the Stack Overflow. 
There are three types of contribution:
\textit{``Accepted Answer"} means that {\tool}'s solution has been adopted as the best answer.  
\textit{``Potential Answer"} means that {\tool}'s solution is listed as a potential answer, but there is no feedback from the question asker yet.
\textit{``Comment"} means that {\tool}'s solution contributes as a comment while other's answer got accepted. 
``QID" refers to question ID  on Stack Overflow, and ``\#ans" refers to the total number of posted answers for the question. 
Overall, {\tool} helps with {\numAns} unanswered issues, and there are 6 solutions that have been accepted as the best answers. 
Fig. \ref{fig:rq4} presents an example of resolved issue posted on Stack Overflow. We can see that, {\tool} can expedite the resolving process of ``BuildProperties" issue by providing a workable solution. 
Specifically, {\tool} can provide the unanswered issues with brief root causes (e.g., 57543742), reference documentation (e.g., 59282213), detailed guidelines (e.g., 66868053), etc. 
{\tool} can also provide timely responses to answer-hungry issues. 
For example, there are 15/26 questions that have no answer at first, and {\tool}'s solutions sever as their only answers {(see red questions with \#ans=1 in Table \ref{tab:so})}.

Summing up, {\tool} helps with unanswered issues on Stack Overflow, and there are 6/{\numAns} solutions that have been accepted as best answers. The results explicitly show that {\tool} can promote knowledge sharing and expedite issue-resolving process.

\section{Discussion and Future Work}

\subsection{Usage Scenario}

\textbf{Serving potential answers}. 
Despite the success of technical Q\&A sites such as Stack Overflow, the answer-hungry problem remains a challenging issue for these forums \cite{DBLP:journals/tosem/GaoXLG21}. 
The mined issue-solution pairs could serve as a knowledge base that can be potentially integrated with, and provide queried information to these Q\&A forums. 
When users ask issues similar to what people have discussed on community chats, the corresponding solutions could be automatically retrieved and recommended as potential answers, relieving the answer-hungry problem to a large extent. 
On the other hand, the extracted issue-solution pairs are also useful for boosting the automation Q\&A in the community live chats by recommending similar questions and corresponding discussions.

\textbf{Boosting developers' profiles}. 
Developers who often provide solutions in community chats may have certain expert knowledge in particular areas. 
According to their historical answers in live chats, the issue-solution pairs extracted can be further used to recommend or assign appropriate respondents to answer questions.
Researchers could also find out which modules or functionalities that the developers are familiar with by analyzing the issue topics that developers have been addressed. Thus, researchers could utilize that information to enhance crowd-sourcing tasks, such as code reviewer recommendation \cite{DBLP:journals/ase/RebaiAMKK20,DBLP:journals/tse/ZanjaniKB16} and issue triage \cite{DBLP:journals/ase/AlmhanaK21,DBLP:journals/tcss/AlazzamALK20}.

\textbf{Highlighting unresolved issues.}
Our approach can effectively find issue-solution pairs from live chats. It can also find
issues where their solutions are empty. Such issues are likely to be the unresolved issues spotted in live chats but are not reported to the project repositories (such as Github).
It is valuable to make the community notice them, e.g., directly push the unresolved issues to the code repository. 
Otherwise, they are likely to be buried in the massive live chats, and the team might miss the opportunity to fix them in time. 
Therefore, a side effect of our approach is to help highlight unresolved issues buried in community live chats.

\textbf{Augmenting Organization/Community Knowledge Base.}
{\tool} can augment the knowledge base of organizations or communities by including discovered issue-solution pairs from group live chats in an automatic and just-in-time way. Moreover, existing technologies such as ontology \cite{2004Handbook} and semantic web \cite{2004A} can be more effective to support information inquiring and sharing across platforms.

\subsection{Where Does {\tool} Perform {Unsatisfactorily}}

\textbf{{Case 1:} Handling dialogs with an issue description utterance lagging behind.} 
{\tool} use the dialog head (all the utterances posted by the dialog initiator before any reply) as the input of our issue model, with the assumption that the dialog initiators are likely to express the issues at the beginning. However, we find that dialog initiators occasionally lag their issue description behind. Here is an example, we can see that, the issue description appears in the $u_{b2}$ utterance, beyond the scope of the dialog head. In such cases, {\tool} cannot accurately detect issues.
{In the future, we plan to dynamically extend the scope of the dialog head, so that the lagging issue descriptions can be included.}

\begin{center}
\vspace{-2ex}
\fbox{\parbox{\columnwidth}{ 
\scriptsize
\noindent
\ttfamily{($u_h$) <A> Good Morning, trying to figure out how to create the DataProvider<DataSetIterator>.  \textcolor{Turquoise3}{[Dialog Head]}\\
($u_{b1}$) <B> Go ahead, no need for greeting!\\
($u_{b2}$) <A> [<-CODE->] incorrect. \textcolor{purple}{How to create the DataProvider from a RecordReaderDataSetIterator? [Issue description]}}
}}
\end{center}

\textbf{Case 2: {Representing }different confidence levels on extracted solutions.}
When discussing issues in community live chats, developers who post the issues often give feedbacks about the solutions provided by their peers at the end. For example, we find that some typical feedbacks are: ``It works.'', ``The issue is fixed.'', ``Figured it out.'', etc. These feedbacks can indicate different confidence levels of the corresponding solution: \textit{Confirmed} and \textit{Candidate}. ``Confirmed" refers to the solution that has been proved to work by the initiator, and ``Candidate" refers to the solution that has the potential to resolve the issue. In the future, we plan to refine the extracted solutions by providing different confidence levels as well. 

\textbf{Case 3: {Differentiating} solutions involving version numbers.}
From the usefulness evaluation results (Table \ref{tab:so}), we notice that the question ``66378139" did not select {\tool}'s solution as its best answer. This is because that the {\tool}'s solution is extracted from the dialog discussing the similar issue in Spring Boot 3.0.0 while the posted issue is related to Spring Boot 2.4.3. Thus, the solution might not be completely suitable. {In the future, we plan to address this issue by linking each issue-solution pair to its corresponding version for better application.}

\textbf{Case 4: {Enhancing} smoothness when combining solution.} 
We directly combine the predicted solution utterances according to their chronological orders as the solution. For the predicted utterances that are not consecutive in the original dialog, logical gaps exist between them. Thus, it may reduce the readability of {\tool}'s solutions. In the future, we would like to improve the readability of the extracted solutions by leveraging language models \cite{2020ERNIE}.

\subsection{Threats to Validity}
The first threat is the generalizability of the proposed approach. It is only evaluated on eight open-source projects, which might not be representative of closed-source projects or other open-source projects. The results may be different if the model is applied to other projects. However, our dataset comes from eight different fields. The variety of projects relatively reduce this threat.

The second threat may come from the results of dialog disentanglement. The accuracy of disentangled dialog has an impact on our results. To reduce the threat, we employed the state-of-the-art technique proposed by Kummerfeld et al. \cite{acl19disentangle}, which outperforms previous studies by achieving 74.9\% precision and 79.7\% recall. Therefore, we believe this can serve as a good foundation for our study on mining issue-solution pairs.

{The third threat relates to the construct of our approach. First, we hypothesize that issue description is likely to appear in dialog head, which is occasionally incorrect in certain cases. 
Second, we do not add version information to issue-solution pairs, which may result in recommending inappropriate solutions. To alleviate the threat, we thoroughly analyzed where our approach performs unsatisfactorily in section VII, and planned future work for improvement.}
Third, the dataset used for the training of the approach includes 750 dialogs and 171 issue-solution pairs from eight Gitter communities, which is not quite large. To avoid the risks of overfitting, we combined dropout with early stopping when training. We observed that the training convergences were achieved at epoch 10-15, and the performance could not be better even more data is given for training.


The fourth threat relates to the suitability of evaluation metrics. We utilize precision, recall, and F1 to evaluate the performance. We use the dialog labels and utterance labels manually labeled as ground truth when calculating the performance metrics. The threats can be largely relieved as all the instances are reviewed with a concluding discussion session to resolve disagreement in labels based on majority voting.

\vspace{-1ex}

\section{Related Work}
\label{sec:related}

\textbf{Knowledge Extraction from Developer Conversations}.
Recently, more and more work has realized that community chat plays an increasingly significant role in software development, and chat messages are a rich and untapped source for valuable information about the software system \cite{chatterjee2019exploratory,DBLP:conf/cscw/LinZSS16,DBLP:conf/msr/ChatterjeeDKP20}. 
There are several studies focusing on extracting knowledge from developer conversations. 
Di Sorbo et al. \cite{DBLP:conf/kbse/SorboPVPCG15} proposed a taxonomy of intentions to classify sentences in developer mailing lists. Huang et al. \cite{Huang2018Automating} addressed the deficiencies of Di Sorbo et al's taxonomy by proposing a convolution neural network (CNN)-based approach. 
Qu et al. \cite{InforSeek_User_Intent_Pred} utilized classic machine learning methods to perform user intent prediction with an average F1 of 0.67. 
Shi et al. \cite{DBLP:conf/icse/ShiXLWL020} proposed an approach to detect feature-request dialogues from developer chat messages via a deep siamese network.  
Rodeghero et al. \cite{DBLP:conf/icse/RodegheroJAM17} presented a technique for automatically extracting information relevant to user stories from recorded conversations.
Chowdhury and Hindle \cite{chowdhury2015mining} filtered out off-topic discussions in programming IRC channels by engaging Stack Overflow discussions as positive examples and YouTube video comments. 
The findings of previous work motivate the work presented in this paper. Our study is different from the previous work as we focus on extracting issue-solution pairs from massive chat messages that would be important and valuable information for OSS developers to check and fix issues.
In addition, our work complements the existing studies on knowledge extraction from developer conversations.

\textbf{Emerging Issue Detection}.
Detecting emerging issues from user feedback timely and precisely is vital for developers to update their applications. Most current work focuses on detecting the emerging issues from short-text social media (e.g., Twitter and Google Play), and determining the emerging issues based on traditional anomaly detection methods. 
For example, 
Guo et al. \cite{DBLP:conf/icse/0002S20} proposed a method for extracting and synthesizing user-reported mini-stories regarding app problems from reviews. 
Vu et al. \cite{DBLP:conf/kbse/VuNPN15} detected emerging issues and trends by counting negative keywords based on Google Play. Since the single words might be ambiguous without contexts, their follow-up work \cite{DBLP:conf/kbse/VuPNN16} proposed a phrase-based clustering approach that relied on manual validation of part-of-speech (PoS) sequences. 
Gao et al. \cite{DBLP:conf/icse/GaoZLK18} presented a topic labeling approach, named IDEA, to automatically detect emerging issues of current versions based on statistics of previous versions. 
Due to the inborn limitations of topic modeling, such as the predefined topic numbers, their follow-up work \cite{DBLP:conf/icse/GaoZD0ZLK19} introduced DIVER which incorporated depth-first pattern mining with version and time-based comparisons.  
Most of these methods focus on detecting emerging issues embedding in short-text social media, while our approach targets to automatically extract issues with their potential solutions (if exists) from community chats, complementing the existing studies on a novel source. In addition, our approach can not only detect emerging issues in community chat, but also extract relevant solutions with resolved issues for reuse purposes, aiming to expedite the issue resolving process.

\textbf{Community-based question and answer extraction}.
Generating large-scale technical question-answer pairs is critical for contributing knowledge that can facilitate software development activities. Existing studies are designed to find questions and corresponding answers from synchronous conversations, i.e., mailing lists and forums. 
Shrestha et al. \cite{DBLP:conf/coling/ShresthaM04} first trained a set of if-then rules to predict questions in email messages, and another set of if-then rules to predict corresponding answers based on features of texts.
Huang et al. \cite{DBLP:conf/ijcai/HuangZY07} presented an approach for extracting high-quality $<$thread-title, reply$>$ pairs from online forums based on SVM classifier and content-quality ranking. 
Cong et al. \cite{DBLP:conf/sigir/CongWLSS08} proposed a sequential pattern-based classification method to detect questions in a forum thread, and a graph-based propagation method to detect answers for questions in the same thread. 
Since previous studies extracted only questions in interrogative forms, Kwong et al. \cite{DBLP:journals/aicom/KwongY12} extended the scope of questions and answer detection, and pairing to encompass also questioned in imperative and declarative forms.
Hen{\ss} et al. \cite{DBLP:conf/icse/HenssMM12} presented an approach to extract FAQs from sources of software development mailing lists automatically.
These approaches utilize the characteristics of their corpora and are best fit for their specific tasks, but they limit each of their corpora and tasks, so they cannot directly transform their methods to the task of extracting issue-solution pairs from community chats.

\vspace{-1ex}
\section{Conclusion}
In this paper, we propose an approach, named {\tool}, to automatically extract issue-solution pairs from development community live chats. {\tool} leverages a novel convolutional neural network by incorporating a basic CNN network with {\attribute} heuristic attributes and \textit{Local-Attention} mechanism to handle the characteristics of this task. 
We build a dataset with 750 dialogs, including 171 issue-solution pairs, and evaluate {\tool} on it. The
evaluation results show that our approach outperforms both issue-detection baselines and solution-extraction baselines by substantial margins. 
By applying {\tool}, we also automatically generate a dataset with over {\numPair} issue-solution pairs extracted from 11 community live chats, and we utilize the dataset to provide solutions for {\numAns} recent issues posted on Stack Overflow.  

\section*{Acknowledgments}
We deeply appreciate anonymous reviewers for their constructive and insightful suggestions towards improving this manuscript.
This work is supported by the National Key Research and Development Program of China under Grant No. 2018YFB1403400, the National Science Foundation of China under Grant No. 61802374, 62002348, 62072442, 614220920020 and Youth Innovation Promotion Association Chinese Academy of Sciences.

\bibliographystyle{IEEEtran}
\bibliography{ref}

\vspace{12pt}

\end{document}